\documentclass{vgtc}                          

\graphicspath{{figures/}{pictures/}{images/}{./}} 

\usepackage{times}                     

\usepackage{tabu}                      
\usepackage{booktabs}                  
\usepackage{lipsum}                    
\usepackage{mwe}                       

\usepackage{mathptmx}                  
\usepackage[normalem]{ulem}

\onlineid{1306}

\vgtccategory{Research}

\vgtcinsertpkg

\providecommand\BibTeX{{\normalfont B{\scshape i\kern-0.0em b}\TeX}}
\usepackage{xcolor}

\definecolor{red}{rgb}{0.9,0,0}
\definecolor{green}{rgb}{0.0,0.7,0}
\definecolor{blue}{rgb}{0.00,0.00,0.75}
\definecolor{orange}{rgb}{0.89,0.58,0.2}
\definecolor{purple}{rgb}{0.6,0.0,0.6}
\definecolor{pink}{rgb}{1,0.03,0.5}
\definecolor{coso}{rgb}{0.2,0.8,0.3}
\definecolor{BonJour}{rgb}{0.7,0.6,0.854}
\definecolor{newblue}{rgb}{0.5,0.00,0.9}

\newcommand{\nothing}[1]{}

\newcommand{\review}[1]{\textcolor{black}{#1}}

\newcommand{\dispT}[1]{true disparity}

\title{Audiovisual Disparities in VR: Impact on Spatial Perception}

\author{%
  Edurne Bernal-Berdun $^\dag$ \thanks{e-mail: edurnebernal@unizar.es} \\ \scriptsize University of Zaragoza, I3A \and
  Mateo Vallejo \thanks{Equal contribution.} \\ \scriptsize University of Zaragoza, I3A \and
  Qi Sun \\ \scriptsize New York University \and
  Ana Serrano \\ \scriptsize University of Zaragoza, I3A \and Diego Gutierrez \\ \scriptsize University of Zaragoza, I3A
}

\teaser{
  \centering
  \includegraphics[width=0.9\linewidth]{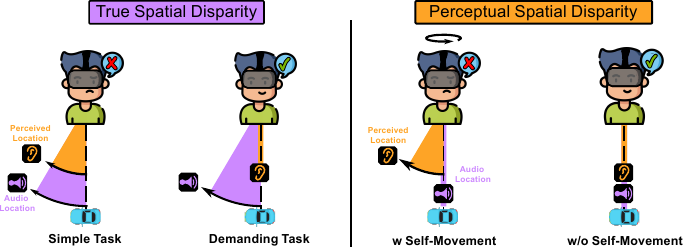}
  \caption{A unified and coherent perception of audiovisual stimuli is crucial for spatial localization in immersive VR experiences. 
  However, such coherence between visual and auditory cues may break down under certain conditions.
  We study two types of spatial audiovisual disparities that can negatively affect the user experience. \textbf{Left:} True disparity arises from an actual misalignment between auditory (purple honk) and visual (blue car) cues. While users are highly sensitive to perceiving slight misalignments (orange), we found that their ability to detect them decreases significantly when focused on engaging tasks \review{inside realistic VR scenes}.
  \textbf{Right:} Perceptual disparity occurs when users perceive a spatial mismatch despite the cues being co-located. We show that while additional environmental cues help mitigate this effect, it persists unless users avoid rotational movements during audiovisual localization. 
  }
  \label{fig:teaser}
}
\abstract{
Virtual reality (VR) experiences often leverage rich and spatialized multimodal environments to increase immersion and engagement. 
This demands a consistent spatial perception of audiovisual stimuli, since perceived discrepancies can disrupt the sense of presence.
In this work, we investigate the consequences of two types of spatial audiovisual disparities: true disparity, where there is a measurable spatial offset between auditory and visual cues, and perceptual disparity, where users report misalignment despite cues being co-located.
Unlike most previous studies that employed controlled but simplified experimental setups, our research focuses on complex, realistic VR environments, allowing us to assess the actual implications for VR content design.
Our experiments indicate that users are highly sensitive to true audiovisual disparities in controlled environments, detecting even minor misalignments. However, when engaged in additional tasks within realistic settings, their ability to notice such discrepancies diminishes significantly. We also observed that previously found perceptual disparities persist in complex audiovisual environments. However, we identify self-initiated head rotations as a key factor; its absence prevents the effect entirely. We hope our findings offer practical insights for designing more immersive and perceptually coherent VR experiences.

} 


\begin{document}



\maketitle
\section{Introduction}

One of the fundamental aspects that contributes to the sense of presence in Virtual Reality (VR) is the spatialization of sensory stimuli~\cite{Hendrix1996sense,bermejo2021relationships,martin2022multimodality}. In the visual domain, 360$^\circ$ rendering ensures that the virtual environment surrounds the user, mimicking real-world experiences. Similarly, spatialized audio positions sound sources in three-dimensional space, enhancing realism and presence. Additionally, proprioceptive feedback, our sense of body position and movement, further reinforces the feeling of being physically present. 
Correct integration of these multimodal sensory cues contributes to the final sense of immersion, creating a coherent and engaging experience~\cite{chandrasekaran2017computational}.


\review{
In the real world, perceptual integration of information from audiovisual sensory inputs facilitates accurate spatial understanding of our surroundings.
However, achieving such multimodal integration has been identified as a core challenge in VR. 
This is due mainly to hardware limitations, such as latency, the lack of focus/accommodation cues from binocular (instead of true 3D) displays, or stereo (instead of spatial) sound. 
As a consequence, audiovisual content in VR applications may result in an erroneous understanding of an object's position. 
While the exact underlying cognitive mechanism remains an open scientific question, such perceptual disruptions do decrease immersion and degrade the user experience~\cite{KimLeeCongruenceEffects}. 
}

Although proper integration of information from different sensory systems enhances immersion, perceptual coherence between modalities is not always ensured.
%
\review{This can occur due to actual misalignments in sensory inputs, which may arise due to hardware limitations, parallax errors in 360$^\circ$ recordings, tracking delays, or inaccuracies in spatial audio rendering~\cite{see2015virtual,Narbutt2018spatialquality}.}
However, even when sensory inputs are perfectly aligned, perceptual aspects can still lead to integration errors. \review{For example, sudden sounds can interfere with the detection of brief visual stimuli~\cite{malpica2020auditory}, and head movements relative to an audiovisual stimulus can cause the auditory cue to be perceived as spatially offset from the visual cue~\cite{Bernal-Berdun2024spatial}}.

Regardless of the cause, such perceptual disruptions can decrease immersion and negatively affect the user experience in VR~\cite{KimLeeCongruenceEffects}. Therefore, understanding proper multimodal integration is essential for identifying situations that result in incoherent perception and ultimately lead to unsatisfactory experiences.

When studying possible inconsistencies in multimodal sensory integration, various aspects of the stimuli, such as temporal, spatial, and semantic factors, are involved. However, the spatial domain, as detailed before, plays a particularly crucial role in VR experiences, as other studies have shown~\cite{kyto2015VEAR,Bernal-Berdun2024spatial}. 
%
Such studies are commonly limited to highly controlled, simplistic setups. While these controlled setups help isolate effects, they fail to capture the complexity of real VR environments, limiting their applicability to VR design. 


We address this limitation by investigating audiovisual disparities (i.e., misalignment between cues) in spatial localization, requiring the integration of auditory, visual, and proprioceptive cues within realistic VR scenes. 
\review{We focus on two potential sources of perceived spatial misalignment between auditory and visual cues in VR. 
The first is a \textit{true audiovisual disparity}, which arises from objective misalignments due to hardware limitations or rendering inaccuracies, such as latency between auditory and visual signals or errors in spatial audio processing \cite{Narbutt2018spatialquality}. 
The second is a \textit{perceptual audiovisual disparity}, where a spatial misalignment between the auditory and visual cues is perceived despite both being physically co-located. This effect arises from the brain’s sensory integration process and can lead to spatial illusions that distort perceived spatial localization.}
%

Our study has three main objectives: first, to determine whether disparities observed in controlled conditions,
\review{either true or perceptual,}
persist in complex environments; second, to identify potential sources of the perceptual disparity; and third, \review{to discuss, based on our findings, strategies to minimize true and perceptual disparities, enhancing perceptual coherence and immersion in VR.} 
\review{To achieve this, we designed and conducted} a series of experiments, described in this paper. For true disparity, we found that while users are highly sensitive to spatial misalignments in controlled environments, their ability to detect them decreases when engaged in a demanding task \review{inside a complex VR scene}. As for perceptual disparity, our results showed that it persists in complex environments even when additional audiovisual stimuli are present to anchor auditory and visual space. However, we also found that its effect, previously attributed to relative stimulus-head velocity, disappears entirely when users remain stationary, and the relative velocity is solely induced by target movement. This points toward self-movement (proprioceptive information) as the principal contributor to the disparity (see Figure~\ref{fig:teaser}).
Based on these findings, we discuss VR design approaches
for minimizing audiovisual spatial disparities, ultimately enhancing immersion and user experience in VR.
\section{Related Work}
Human perception relies on multiple sensory modalities, including vision, audition, somatosensation, and vestibular input, which are integrated through multimodal integration. This process is not a simple aggregation of sensory inputs but a dynamic integration of modalities based on task demands. In neuroscience and psychology, this challenge is known as the Neural Binding Problem~\cite{Feldman2013NeuralBinding}, the process by which the brain combines different sensory signals into a unified percept. Due to this multimodal nature, interactions between sensory modalities can enhance perception but may also lead to misattributed bindings, where unrelated stimuli are perceived as a single event or a single event is attributed to unrelated stimuli, giving rise to cross-modal illusions.
 
Audiovisual disparities often occur in the temporal domain, with well-documented illusions demonstrating cross-modal influence. The McGurk effect~\cite{mcgurk1976HearingLips} illustrates how visual input (i.e., vocalization) can alter perceived speech sounds, even when auditory and visual cues conflict, highlighting Visual Dominance~\cite{posner1976visualdominance}, which describes how vision is usually the sense that takes most weight during multimodal integration. 
However, auditory stimuli can also influence vision. \review{For instance, the auditory system’s greater temporal precision can lead it to dominate perception in certain contexts.} 
The double flash illusion~\cite{shams2000what} shows that a single visual flash paired with two beeps can create the illusion of two flashes. In contrast, Malpica et al.~\cite{malpica2020auditory} found that a brief sound presented alongside a visual stimulus can cause the visual input to go unnoticed entirely. Further studies~\cite{shams2010crossmodal} reveal that sound can also modulate perceived brightness, duration, and motion of visual stimuli.

Audiovisual disparities also arise in the spatial domain, affecting localization, that is, the perception of a stimulus’s position relative to the observer and the environment. Visual and auditory localization rely on distinct sensory mechanisms with differing accuracies. Humans can localize visual stimuli with as much accuracy as their visual acuity ($\approx$ 5 arc minutes or 0.00023$^\circ$)~\cite{snellen1873probebuchstaben}, whereas auditory spatial discrimination is rougher, typically around 1$^\circ$–2$^\circ$ in optimal conditions~\cite{Perrott1990MAA,Grantham2003MAA}. While vision generally dominates spatial localization due to Visual Dominance, auditory cues can take precedence when visual reliability decreases~\cite{Gao2020CurvGains}. A well-known example of cross-modal influence is the ventriloquist effect~\cite{hairston2003VisualLocalization}, where a mismatched sound is perceived as originating from a visual source, \review{illustrating how sensory integration can occur despite the audiovisual spatial disparity.}
%
Prior research has examined the limits of spatial disparity in audiovisual binding. Kyto et al.~\cite{kyto2015VEAR} found that in Augmented Reality, the ventriloquist effect persisted for spatial incongruences up to 30°. \review{Kim and Lee~\cite{KimLeeCongruenceEffects} investigated various audiovisual incongruences in Virtual Reality, including temporal, spatial, and semantic misalignments, while analyzing their impact on engagement and immersion.}

However, audiovisual spatial localization is typically performed alongside head movements, engaging both the proprioceptive and vestibular systems. The vestibular system, located in the inner ear, plays a crucial role in sensory integration, as seen in the vestibulo-ocular reflex, where head movements trigger compensatory eye movements to stabilize gaze~\cite{angelaki2004eyes}. Consequently, both vision and hearing are influenced by self-motion.
Ross et al.~\cite{ross1997compression} observed that rapid eye movements induce perceptual compression of visual space, reducing its perceived horizontal extent. Similarly, Leung et al.~\cite{leung2008compression} reported compression of auditory space during head rotations, indicating that movement affects spatial localization. Genzel et al.~\cite{Genzel2018sounddepth} found that head movements enhance localization accuracy, but only when self-initiated, suggesting that the perception of self-motion contributes to spatial understanding. Bernal-Berdun et al.~\cite{Bernal-Berdun2024spatial} further revealed that head-body rotations can cause a systematic overestimation of the location of the auditory stimulus in the direction of movement in VR.
This work builds on this finding, investigating the underlying factors of this perceptual overestimation and proposing strategies to mitigate its impact. 

While prior research has assessed audiovisual localization accuracy and identified perceptual limitations, its implications for practical VR environment design remain unclear. Therefore, we investigate the actual impact of both true and perceptual disparities on complex VR experiences and discuss design guidelines to mitigate their potential effects.

\section{Impact of True Audiovisual Disparities}
\label{sec:true_disparities}
\begin{figure}[t]
	\centering
	\includegraphics[width=\columnwidth]{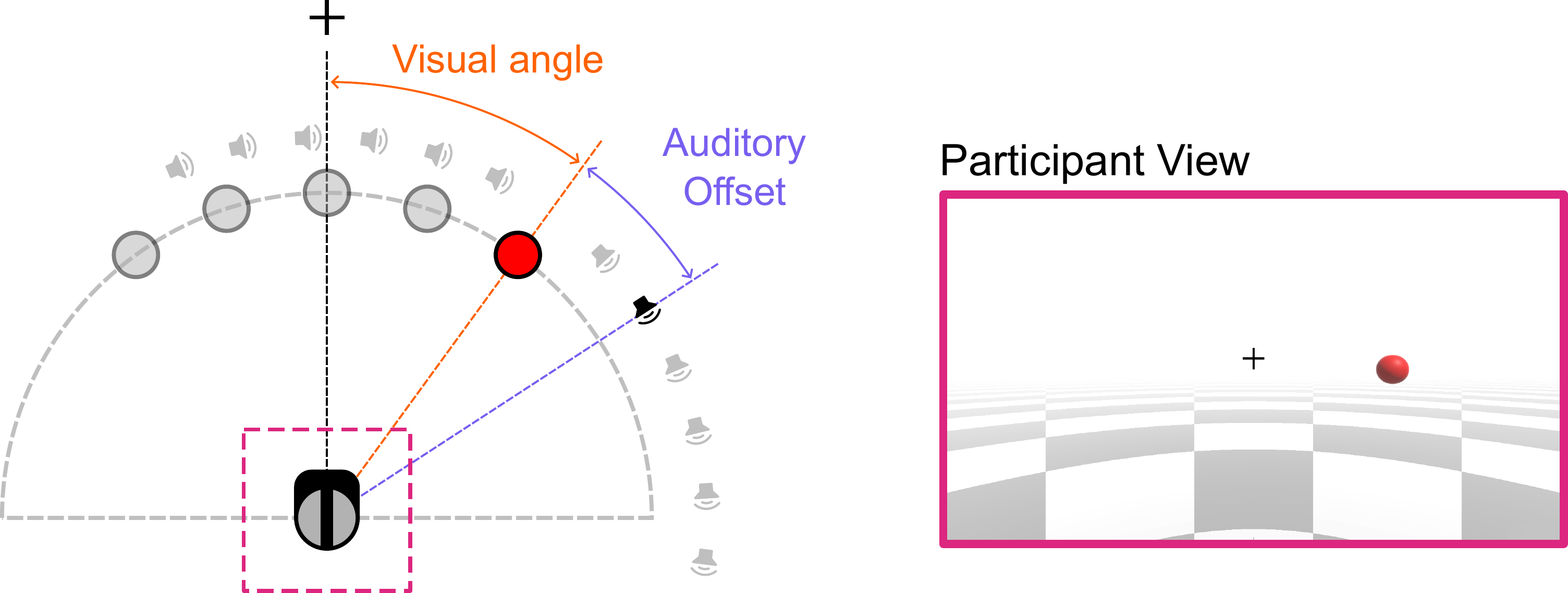}\caption{Baseline experiment setup: Participants are shown a visual cue, a red ball, that appears at one of five possible locations in front of them ($\pm$36º, $\pm$18º, or 0º). An associated auditory cue is either co-located with the visual cue (0º audiovisual disparity) or shifted by $\pm$10º, $\pm$20º, $\pm$30º, $\pm$40º, $\pm$50º, or $\pm$60º. Participants complete a 2AFC task, indicating whether they perceive the auditory cue to the left or right of the visual cue.}

\label{fig:congitiveBaselineSetup}
\end{figure}

\begin{table}[h]
    \centering
    \caption{Obtained PSE, leftward detection threshold (DT$_{25\%}$), and rightward detection threshold (DT$_{75\%}$) from the psychometric curves fitted for both the baseline and the demanding task conditions. The confusion region, defined as the difference between the detection thresholds, is denoted as CR$_{25\%-75\%}$.}
    \begin{tabular}{lcccc}
 Condition & DT$_{25\%}$ & PSE    & DT$_{75\%}$ & CR$_{25\%-75\%}$ \\ \midrule \midrule
 Baseline &  -4.61º & -0.40º & 3.82º & 8.43º\\
 Task demand & -26.34º  & -8.00º & 10.33º & 36.67º\\
    \end{tabular}
    \label{tab:PSE_DT_values}
\end{table}

In this section, we study the impact of spatial \emph{true disparities}, where the auditory and visual cues are misaligned in space.
If users perceive this misalignment, it can disrupt immersion or cause confusion in VR. However, previous research suggests that even when a misalignment exists, both cues can still be integrated into a unified percept as long as the disparity does not exceed a certain threshold~\cite{jack1973effects}. Therefore, identifying these thresholds may help designers to create environments that avoid perceptual conflicts. 
\review{Studying these thresholds in complex, realistic scenarios is essential, offering a deeper insight into real-world user experience; as such, thresholds may differ from those observed in simplified, controlled environments.}

For this purpose, we conducted two different experiments to study users' detection thresholds. First, we measured the threshold at which users detect a \dispT{} in a controlled, simplified setting, where a red ball emits a white sound with varying levels of misalignment between the visual and auditory cues. This experiment allowed us to estimate participants' sensitivity to spatial misalignment in VR. Then,  we measured the same thresholds in a more complex, realistic scenario, where participants performed a visual search task (finding figurines) within an immersive environment presenting two audiovisual stimuli with misaligned cues. By comparing results across conditions, we assess whether the thresholds remain stable or shift under more demanding tasks and complex environments.

Similar to previous work~\cite{kyto2015VEAR}, we used the Two-Alternative Forced Choice (2AFC) method to assess participants' ability to detect audiovisual true disparities. It consists of presenting audiovisual stimuli to participants where the audio is misaligned to the left or right of the visual cue, and asking them whether they perceived the audio as positioned to the right or to the left of the visual cue. 
We then calculated the proportion of rightward responses (i.e., instances where participants perceived the audio as coming from the right) across different auditory offsets. A psychometric function is then fitted to estimate participants' detection threshold (DT) for audiovisual misalignments~\cite{SCHUTT2016BayesianPsy}. The detection thresholds for rightward and leftward misalignments correspond to 75$\%$ and 25$\%$ of rightward responses, respectively. Additionally, this psychometric fitting reveals the point of subjective equality (PSE), which indicates the position at which the auditory cue is perceived as perfectly aligned with the visual cue.

The experiments were performed with an HTC Vive Pro Eye head-mounted display. \review{Directional audio was reproduced through the built-in headphones of the headset, using the Steam Audio SDK, which employs a generic Head-Related Transfer Function (HRTF) with a resolution of 5$^{\circ}$ in the horizontal plane.}
Additional hardware details and specifications can be found in the supplementary material.
The research protocol was approved by \textit{the Comité de Ética de la Investigación de la Comunidad Autónoma de Aragón} (CEICA).

\begin{figure}[t]
	\centering
	\includegraphics[width=\columnwidth]{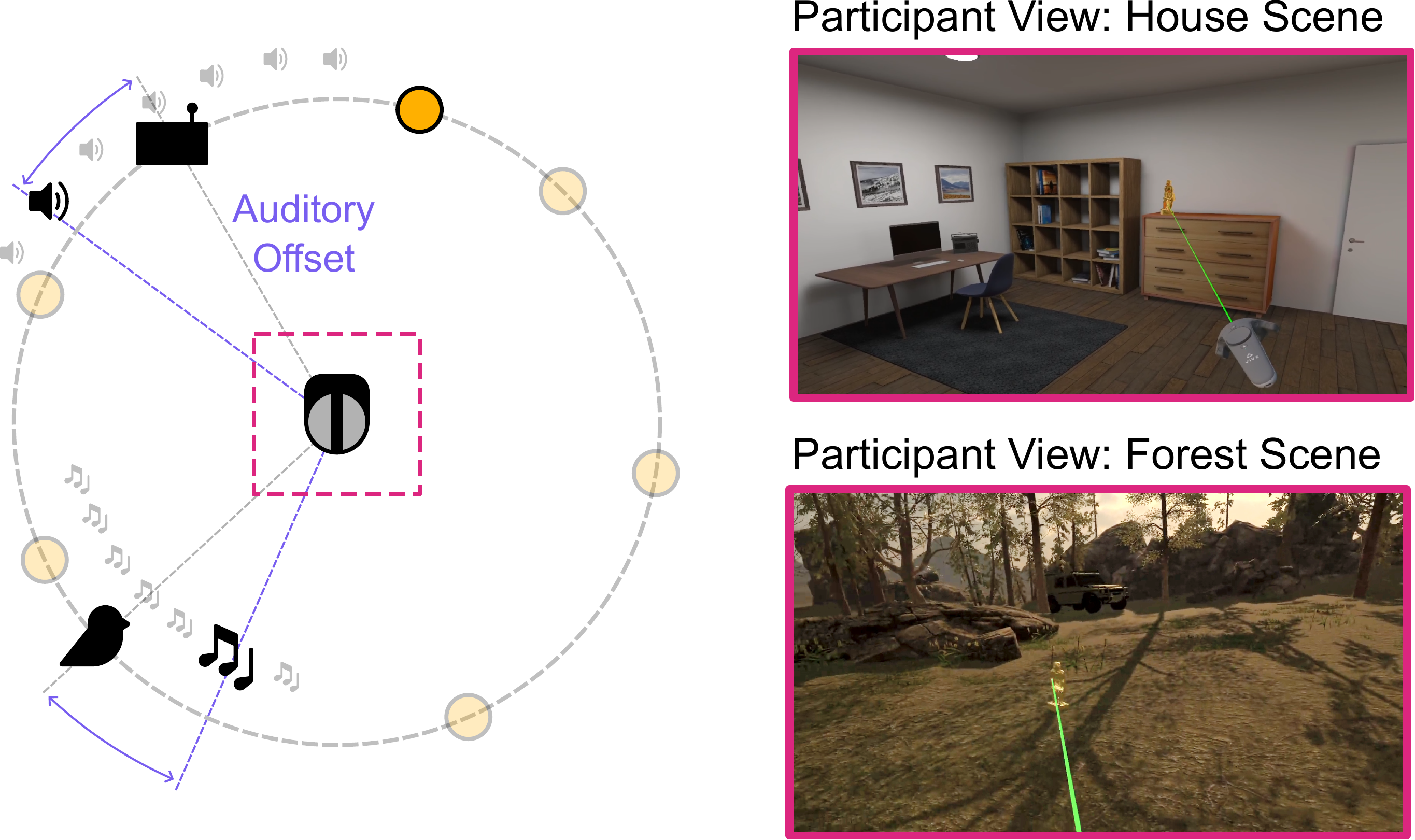}
	\caption{Task demand experiment setup: Participants are immersed in two different scenes, where two random audiovisual stimuli are presented in each (e.g., bird and radio in the figure). They complete two tasks: first, they search for as many figurines as possible (yellow dots); second, they perform the same 2AFC task as in the baseline experiment, indicating whether they perceive the auditory cues to the left or right of the visual cues.}
\label{fig:congitiveEnvironmentSetup}
\end{figure}

\begin{figure*}[t]
	\centering
	\includegraphics[width=\textwidth]{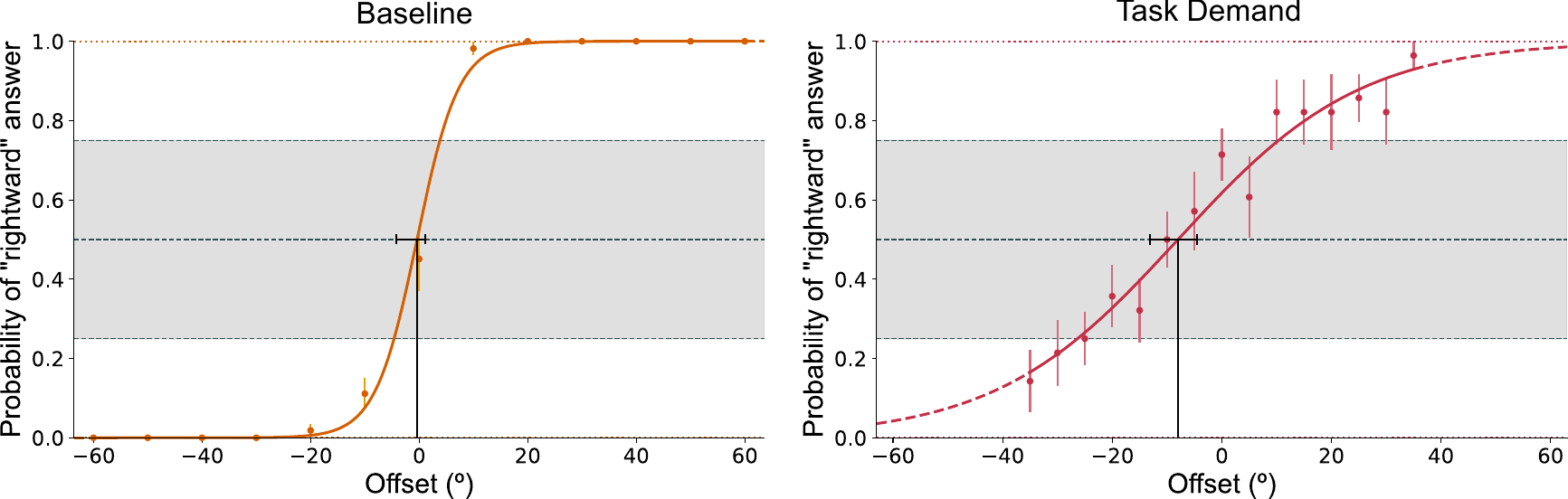}
	\caption{Fitted psychometric curves for the baseline (\textbf{left}) and task demand (\textbf{right}) experiments. The x-axis shows the offset in degrees applied to the auditory cue relative to the visual cue (positioned at 0$^\circ$). The y-axis indicates the proportion of trials in which participants perceived the auditory cue as coming from the right of the visual cue for a given offset, which can be applied either to the left (negative offset values) or to the right (positive offset values) of the visual cue. Each point represents the participants' pooled responses, with error bars showing the standard error. The PSE, DT${25\%}$, and DT${75\%}$ are indicated by dashed gray horizontal lines, and the confusion region is shaded in gray. The curves show how the offsets required to detect misalignments notably increase when participants perform an additional task.}
    \vspace{-8pt}
\label{fig:PsychometricCurves}
\end{figure*}

\subsection{Baseline Detection Accuracy}
\label{sec:baseline}

We first assessed participants' accuracy in detecting true disparities in a simple and controlled environment. Their task was to determine whether a misalignment existed between auditory and visual cues in an audiovisual stimulus: a red ball emitting white noise. The stimulus appeared at various angular positions relative to the participant’s forward direction.
Each trial presented the stimulus at a specific location with an auditory offset. After two seconds, the stimulus disappeared, and participants indicated whether they perceived the auditory cue to the right of the visual cue. They were allowed to move their heads to align with the stimulus before responding.
Before each trial, we realigned participants by having them fixate on a cross indicating the forward direction.

\emph{\textbf{Technical Details:}} The environment featured a white background and a polar checkerboard-patterned ground, with a black cross indicating the forward direction (see Figure~\ref{fig:congitiveBaselineSetup}). The audiovisual stimulus consisted of a red ball (visual cue) and white noise (auditory cue). It appeared for two seconds at a random position among $\pm36^\circ$, $\pm18^\circ$, and $0^\circ$~\cite{kyto2015VEAR}, 3 meters away from the participant. \review{The two-second duration was established to ensure that participants had sufficient time to accurately identify the source of the sound~\cite{Bernal-Berdun2024spatial}}. \review{Following previous works that consider angular offsets in the range of $5^\circ$ to $10^\circ$~\cite{kyto2015VEAR,KimLeeCongruenceEffects}, our }auditory cue could be offset from the visual cue by $\pm10^\circ$ to $\pm60^\circ$ in $10^\circ$ increments or co-located (offset = $0^\circ$). Participants completed 65 trials, covering the five different locations and 13 possible offsets. 

\emph{\textbf{Participants:}} We performed this experiment with 13 participants (11 male and 2 female). All participants reported having corrected-to-normal hearing and vision. Two participants were discarded due to presenting random behavior, not being able to discern the disparity under the extreme $\pm60^\circ$ tested offsets, acting as sentinel trials. 

\subsubsection{Results}
To fit the psychometric curve, we pooled the responses of all participants for the different offsets of the auditory cue, independently of the position of the ball, as we did not observe a significant effect of the visual cue position in the results (see supplementary material). The psychometric curve, PSE, and DTs values are shown in Figure~\ref{fig:PsychometricCurves} and Table~\ref{tab:PSE_DT_values}. The PSE is almost perfectly centered at $0^\circ$ (PSE = $-0.40^\circ$), indicating that participants correctly perceive auditory and visual cues as aligned when they are indeed aligned. The DTs, where the auditory cue is perceived to the left (DT$_{25\%} = -4.61^\circ$) or right (DT$_{75\%} = 3.82^\circ$), are lower than $5^\circ$, matching the measured horizontal spatial resolution of the HRTF used to render the audio. The same trends can be observed in the individual psychometric fits for each participant, which are included in the supplementary material. This suggests that participants exhibit almost perfect accuracy in detecting true audiovisual disparities under ideal conditions, making slight misalignments highly noticeable.
These DTs are more restrictive than those reported by Kyto et al.~\cite{kyto2015VEAR} for AR and Kim and Lee~\cite{KimLeeCongruenceEffects} for VR, where misalignments between the cues of $15^{\circ}$ or less were often perceived as a unified audiovisual stimulus. The difference may result from methodological variations such as stimulus duration. We presented stimuli for two seconds while Kim and Lee~\cite{KimLeeCongruenceEffects} used only 100 milliseconds. Hardware differences may also play a role, as depth perception in AR is typically less accurate than in VR, potentially affecting spatial judgment. 

Our results show that participants reliably detect small misalignments when they have at least two seconds to make a judgment and can align themselves with the audiovisual stimulus. This highlights a critical concern for VR content: even slight audiovisual misalignments can be noticeable and may impact the user experience.

\subsection{Influence of Task Demand}
\label{sec:task_demand}

In this experiment, we assess whether the sensitivity to slight true disparities observed in the baseline experiment persists in a richer VR environment, more representative of real applications. While participants could easily detect auditory misalignments, the impact on VR experiences where users are engaged in different tasks remains unknown. To explore this, we measured detection thresholds for audiovisual true disparities in more complex VR scenes (a forest and a house) while introducing an additional task, a visual search. Participants are tasked with identifying spatial 
 misalignments
between auditory and visual cues while searching for hidden figurines within rich audiovisual environments. These environments always presented two clear audiovisual stimuli that produced sound during the whole trial. Before the experiment, participants were informed that figurines would appear randomly and that auditory and visual cues might have spatial offsets. \review{At the end of each trial, once they could no longer search for figurines,} they answered the same 2AFC question as in the baseline experiment, indicating whether the auditory cue was to the right or to the left of the visual cue.

\emph{\textbf{Technical Details:}} \review{Each 15-second trial involved searching for figurines that appeared consecutively while audiovisual stimuli, with misaligned auditory cues, were present in the scene. At the end of each trial, participants answered the 2AFC question}. Two of five possible audiovisual stimuli for each scene are randomly selected and remain visible with their sounds playing, while the rest stay hidden and muted (see Figure~\ref{fig:congitiveEnvironmentSetup}). \review{No sound sources other than the audiovisual stimuli were included in the scenes.} In each trial, both auditory cues are shifted by an offset between $-30^\circ$ to $+30^\circ$ in $5^\circ$ increments. All offset values are tested across two scenes, totaling 26 trials per participant. The trial order is randomized, causing participants to switch between different offsets and scenes throughout the experiment.
The two scenes presented different audiovisual stimuli: a \textit{forest}, featuring a waterfall, a radio, a car, a crackling bonfire, and howling wolves; and a \textit{house}, with stimuli including a washing machine, a radio, a running kitchen faucet, a singing bird by the window, and a ticking clock. The audiovisual stimuli were selected for their clear sound-source association and relevance to the environment. They were strategically placed to ensure they were evenly distributed, preventing more than two from being visible at once\review{, and ensuring all stimuli were equally audible (further details can be found in the supplementary material).}

\emph{\textbf{Participants:}} We performed this experiment with 14 participants (11 male and 3 female). All participants reported normal or corrected-to-normal vision and audition.

\subsubsection{Results}
\label{sec:results_task_demand}
\review{Participants successfully engaged with the additional task, finding 2.231 figures ($\sigma=0.273$) per each 15-second trial, with a mean discovery time of 4.025 seconds per figure ($\sigma=0.718$).} As we did not find any significant differences between forest and house scenes, we pooled participants’ responses across both scene types for the psychometric fitting shown in Figure~\ref{fig:PsychometricCurves}. We include the comparison between scenes and individual psychometric fittings in the supplementary material. \review{While the selected scenes do not capture the full range of visual and auditory complexity, they are sufficient for determining whether increased complexity, combined with an additional task, influences the detection of the true disparity.} The fitted psychometric curve, reveals a larger confusion area between DTs ($36.67^\circ$) compared to the baseline experiment ($8.197^\circ$), indicating significantly lower sensitivity to true audiovisual disparities (see Table~\ref{tab:PSE_DT_values}). Additionally, the standard error (shown as vertical color bars in Figure~\ref{fig:PsychometricCurves}), calculated from individual participant percentages, increases significantly compared to the baseline condition.
\review{An independent samples t-test on PSEs obtained from individual psychometric fits per participant revealed a significant difference between the baseline ($\mu = -0.72^\circ$, $SE = 0.83^\circ$) and task demand conditions ($\mu = -7.21^\circ$, $SE = 1.57^\circ$), $t(23) = 3.60$, $p<0.05$, $r = 0.60$. Assumptions of normality (Shapiro–Wilk: $W_{baseline} = 0.870$, $p = 0.076$; $W_{task} = 0.976$, $p = 0.944$) and homogeneity of variances (Levene’s test: $F(1, 23) = 3.82$, $p = 0.063$) were met. 
We additionally compare the individual confusion regions CR$_{25\%-75\%}$. Due to violation of homogeneity of variances (Levene’s test: $F(1, 23) = 23.213$, $p < 0.05$), a non-parametric Mann–Whitney U test was conducted. Results revealed a significant difference in CR$_{25\%-75\%}$ between the baseline ($\mu = -8.34^\circ$, $SE = 0.72^\circ$) and task demand conditions ($\mu = -28.79^\circ$, $SE = 4.48^\circ$), $U = 130.00$, $p < 0.05$, with a large effect size, $r = 0.59$.}

These results suggest that the inclusion of a more complex environment, where participants perform an additional demanding task, reduces their sensitivity to true audiovisual disparities. \review{Although whether this is due solely to the added task or also to several aspects associated with the increased audiovisual complexity remains an open question.} The PSE is also shifted leftward (PSE~$=$~$ -8.00^\circ$), suggesting that participants tend to perceive the auditory stimulus slightly shifted to the right of their actual location. This rightward bias in perception is consistent across the two environments and aligns with previous studies showing a natural rightward bias in human auditory and visual perception~\cite{Dufour2007Rightward,SOSA2010rightbias}, which tends to become more pronounced as perceptual load increases~\cite{chen2017hemispheric,naert2018asymmetric}.

Nevertheless, the overall increase in the detection thresholds suggests that when users are engaged in tasks unrelated to detecting misalignments and the environment is more complex, they can tolerate bigger true audiovisual disparities without noticing them.

\section{Impact on Perceptual Audiovisual Disparities}
\label{sec:perceptual_disp}

The previous section studied audiovisual true disparities produced by an actual misalignment between the visual and auditory cues. However, even when the sensory information is accurate, the multimodal integration of this information may be incorrect, producing a \emph{perceptual audiovisual disparity}. 

Therefore, we further explore the underlying causes of the spatial perceptual disparity identified by Bernal-Berdun et al.~\cite{Bernal-Berdun2024spatial}.
Their study found that in VR, users misperceive the auditory cue location of audiovisual stimuli toward the direction of the head-body rotation, attributing this to the relative velocity between the head and the stimuli. They showed its negative impact through a VR application and proposed a compensation technique.
In their study, participants identified which of four possible visual targets was emitting a spatialized sound. The study found that when participants performed the task while actively rotating, they tended to perceive the location of the sound as being offset in the rotation direction, consistently misidentifying the target as the one next to the real target.
To address this, they proposed shifting the sound source in the opposite direction to compensate for the perceptual disparity. While effective, this method is challenging to implement in VR, as it requires detecting relative motion between the head and stimulus and dynamically adjusting the offset based on head rotation speed. To develop more practical mitigation strategies, we aim to better understand the underlying causes of this perceptual disparity.

We first replicate their experimental conditions in a richer, realistic VR environment. Their study used only singular auditory cues, but real-world applications often involve more complex soundscapes. We then study whether additional audiovisual stimuli help anchor the auditory and visual reference systems or if perceptual disparity persists even in richer environments (Section \ref{sec:environmental}). Secondly, we explore how self-movement, which introduces proprioceptive information to the integration process, impacts the perceptual disparity. To isolate this effect, we introduce relative velocity exclusively through target motion, decoupling it from head rotation (Section \ref{sec:selfMovement}).

\review{The experiments were performed with the HTC Vive Pro Eye HDM, reproducing the sound through the built-in headphones using the same generic HRTF as in the previous experiments~(see Section~\ref{sec:true_disparities}); further hardware details can be found in the supplementary material.}
The research protocol was approved by \textit{the Comité de Ética de la Investigación de la Comunidad Autónoma de Aragón} (CEICA).

\subsection{Environmental Cues}
\label{sec:environmental}
Existing experiments study an audiovisual environment with only a single auditory cue. This contrasts with real-world applications, where auditory environments are typically richer and spatially contextualized. We investigate whether the lack of additional audiovisual stimuli impairs localization, as previous research suggests that auditory localization benefits from parallax~\cite{Genzel2018sounddepth}, or if the perceptual disparity remains unaffected.

\subsubsection{Experimental Procedure}

The general structure of the experiment mimics the Bernal-Berdun et al.~\cite{Bernal-Berdun2024spatial} study, with an additional focus on engaging participants by making the experience more immersive. The environment consisted of a realistic forest scene with dynamic vegetation, and the experiment was presented as an archery game.
Participants selected the targets by firing an arrow with a bow, for which they needed both hands to operate.
The bow included a mechanism that displayed the trajectory the arrow would take once fired to remove the possibility of participants accidentally selecting a different target.
To evaluate the influence of additional audiovisual stimuli, we introduced two conditions: including or not including additional audiovisual environmental cues.

\begin{figure}[t]
	\centering
	\includegraphics[width=\columnwidth]{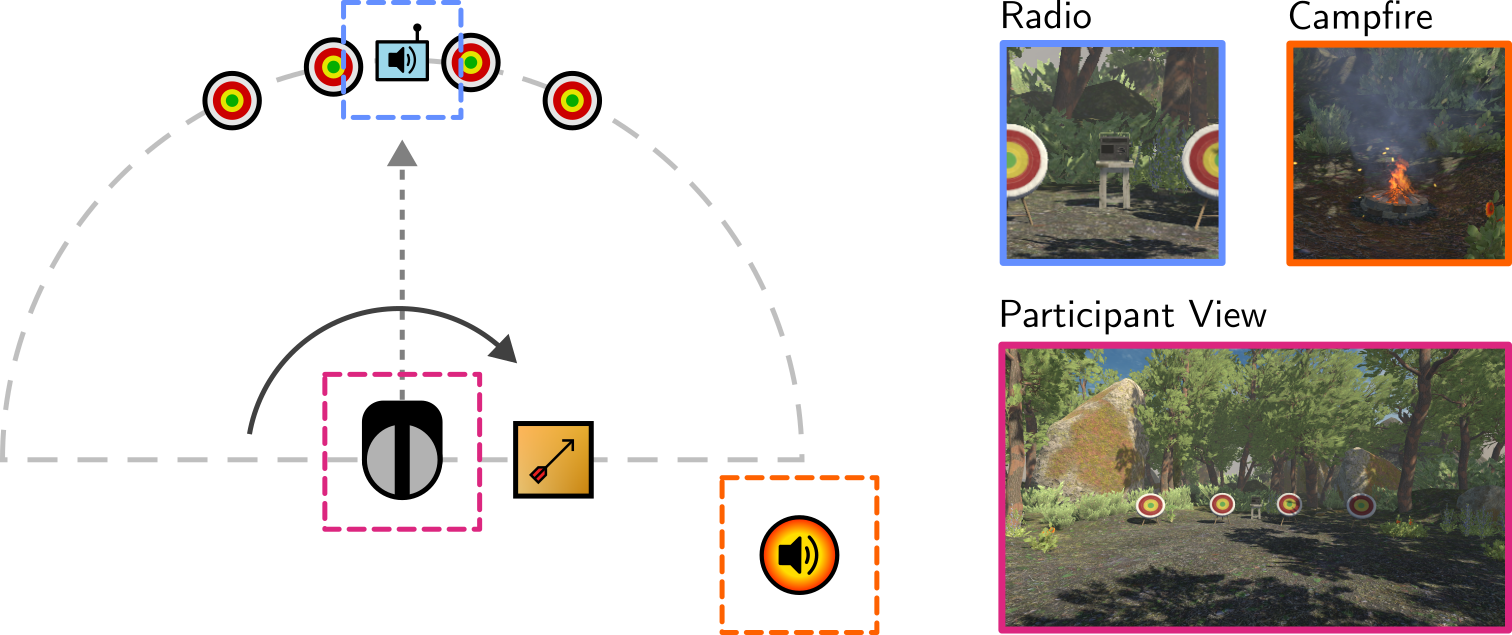}
	\caption{Environmental cues experiment setup. Participants first identify which of the four targets emits a sound while performing a 180$^\circ$ rightward rotation, and then select the target they identified as having emitted the sound. Participants perform this task in two conditions: with and without the additional environmental audiovisual stimuli (radio and campfire).}
	\label{fig:environmentSetup}
    \vspace{-8pt}

\end{figure}

\emph{\textbf{Technical Details:}}
The possible visual cues, presented as archery targets, had $5^\circ$ of visual angle size and were placed at a distance of 10 meters from the participant, with a separation of $15^\circ$. The auditory cue, only present during rotation, reproduced a clockwork mechanism sound. Participants were instructed to rotate 180$^\circ$ to grab a new arrow while focusing on the audiovisual stimulus among potential targets. After completing the rotation, the auditory cue was muted, and they were asked to fire at the target they perceived as co-located with the auditory cue. Two different conditions were presented: \textit{with} and \textit{without} the additional environmental stimuli. The order in which participants perform each modality \review{was} selected at random. In the environmental stimuli condition, two stimuli were added to the scene: a radio, located between the central targets, and a campfire, located at the ending position of the participant's rotation (see Figure~\ref{fig:environmentSetup}).
During each condition, each of the four targets (i.e., possible visual cues) is sampled five times, resulting in 20 trials per condition, amounting to 40 trials per participant. Four practice trials were performed at the beginning to help participants familiarize themselves with the task.

\emph{\textbf{Participants:}}
We performed the experiment with 19 participants (12 female and 7 male). All participants reported having corrected-to-normal hearing and vision.
One participant was discarded due to random guess behavior. Participants were not informed of the hypothesis of the study.

\subsubsection{Results}
\begin{figure}[t]
	\centering
	\includegraphics[width=\columnwidth]{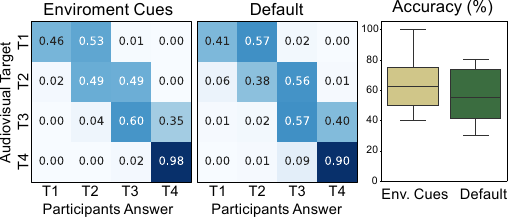}
	\caption{\textbf{Left:} Accuracy confusion matrix for target selection in conditions with and without environmental cues. Higher diagonal accuracy indicates improved identification when environmental cues are present. \textbf{Right}: Overall participant accuracy in both conditions, with and without (default) environmental cues.}
    \vspace{-4pt}
\label{fig:environmentConfusion}
\end{figure}

Participants’ mean head rotation speed during the experiment was $112.42^\circ/s$, less than $3\%$ different from prior literature~\cite{Bernal-Berdun2024spatial}, suggesting consistency in participants' behavior with their study.
We compared localization accuracy between conditions with and without environmental audiovisual stimuli. While median accuracy was slightly higher \textit{with} ($63.2\%$) than \textit{without} ($56.4\%$), confusion matrices (Figure~\ref{fig:environmentConfusion}) show a consistent pattern between conditions \review{(further details can be found in the supplementary material)}. Participants usually misidentify the true visual cue, selecting targets adjacent to the sound-emitting target in the direction of rotation (shown by a displacement of the diagonal values).
The difference between conditions' accuracy distributions satisfies the normality assumption (Shapiro-Wilk test, w$=$0.96, p$=$0.65). Therefore, we conducted a Student’s t-test, which revealed that participant performance with environmental stimuli ($\mu$=63.2$\%$, SE$=$0.039) was not significantly higher than without environmental stimuli ($\mu$=56.4$\%$, SE$=$0.039), t(17)$=$1.998, p$=$0.10. 
%

These results indicate that perceptual disparity persists regardless of environmental audiovisual anchors. Although the inclusion of stimuli slightly reduced errors, the disparity and its associated problems remain in more realistic environments.

\subsection{Implications of self-movement}
\label{sec:selfMovement}

In the previous section, we observed that the perceptual disparity arises when there is relative motion between the head and the audiovisual target. Since this disparity and its related issues seem to persist across environments, independently of the number of audiovisual stimuli, we now seek to explore its underlying cause. Specifically, we investigate whether self-movement plays a fundamental role in this effect, given its known influence on perception~\cite{panerai2002contribution,Serrano2020LateralMotion}. We seek to determine whether the discrepancy arises solely from relative motion or if self-movement (e.g., proprioceptive cue) is a key factor.
To examine the impact of the proprioceptive and vestibular systems, we designed an experiment where relative motion between the head and target was preserved, but participants remained stationary while the targets moved instead. 
Additionally, we applied the enhanced target localization method proposed by Bernal-Berdun et. al~\cite{Bernal-Berdun2024spatial} which consisted of offsetting the sound source according to their measured PSE, which represents the angle where auditory and visual cues are perceived as aligned. 
\review{This condition allowed us to assess whether the perceptual disparity persists. If the disparity is caused by head-stimulus relative motion, performance should improve with compensation. Contrarily, if no disparity is present, this condition could reveal potential negative effects of applying the previously proposed compensation.}
Understanding whether this perceptual issue is specific to self-movement will help define the conditions under which it becomes problematic.

\subsubsection{Experimental Procedure} 
\begin{figure}[t]
	\centering
	\includegraphics[width=\columnwidth]{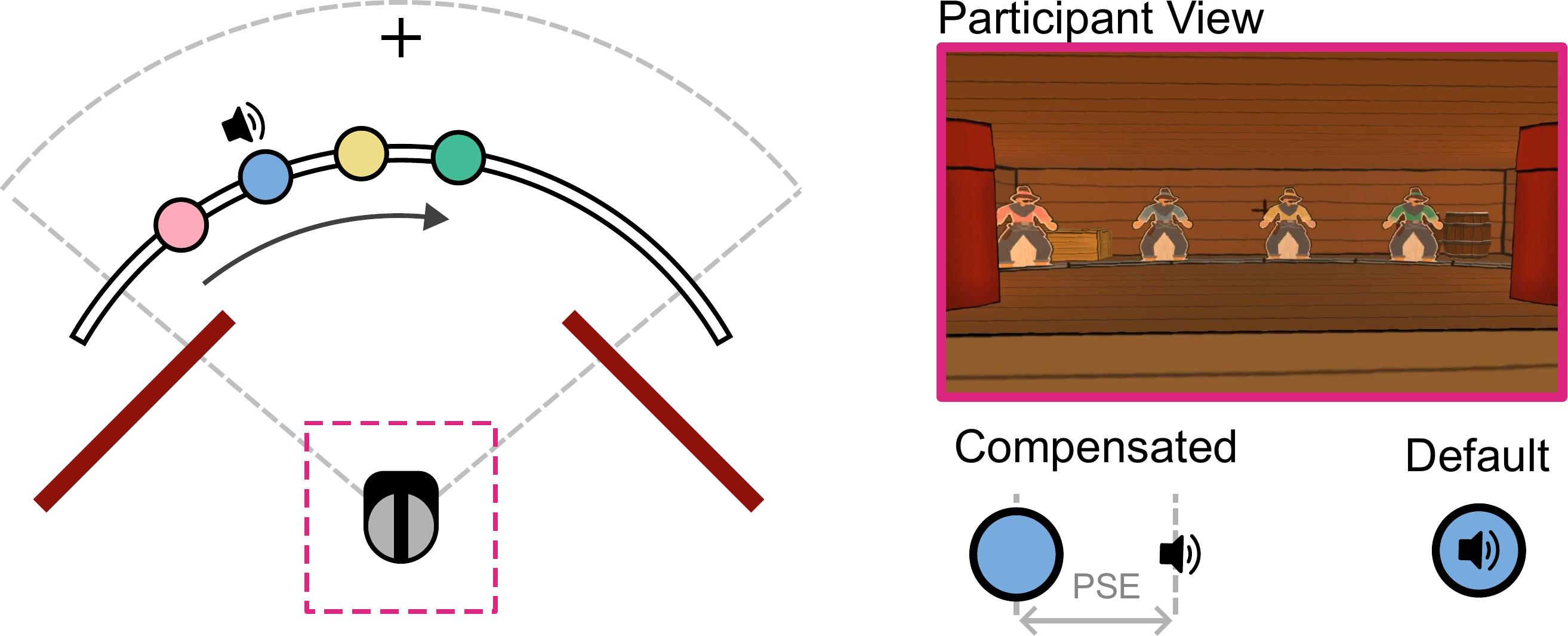}
	\caption{Experiment setup. Participants first identify which of the four moving targets emits a sound as they travel from left to right. Then, the targets reappear and participants select the target they perceived as having emitted the sound.}
	\label{fig:targetMovingSetup}
\end{figure}

We sought to replicate the same experiment as the previous section while eliminating participant movement. We presented the experiment as a carnival shooting gallery, where the four possible visual targets move along a rail across the room while one of them emits a sound  (see Figure~\ref{fig:targetMovingSetup}). At the end of the movement, participants are asked to identify the \textit{audiovisual target} by determining the sound’s \review{location}. They are instructed to face forward and avoid head rotations. Once the targets leave the participant’s field of view, the target's sound ceases, and they reappear at the center, enabling participants to select the one they perceived aligned with the auditory cue. 
We tested two conditions based on the previous work experiment: \textit{default}, where the auditory cue remains perfectly aligned with the visual cue, and \textit{compensated}, where the auditory cue is shifted in the opposite direction of movement by an amount equal to the previously measured PSE. Each participant completed 16 trials, two per condition and visual target, randomly shuffled. Additionally, four practice trials with a randomly assigned target and no audio offset were conducted initially to help participants familiarize themselves with the task.

\emph{\textbf{Technical Details:}} Consistent with the previous work experiment, the targets moved at a constant speed of $50^{\circ}/s$. They followed an arc-shaped trajectory from left to right at a 10-meter distance from the participant, with $15^{\circ}$ separation between them, and a height of $5^{\circ}$. From the participant’s perspective, this motion was equivalent to a leftward head rotation. When compensating for the perceptual disparity, the target's audio cue is shifted by the leftward PSE ($+9.52^{\circ}$)~\cite{Bernal-Berdun2024spatial}.
If participants rotated their heads by more than $10^{\circ}$ in any direction during the trial, it was interrupted and shuffled back into the trials pool. This restriction ensured that participants maintain a forward-facing position.

\emph{\textbf{Participants:}} The experiment was conducted with 17 participants (11 female and six male), all of whom reported having corrected-to-normal hearing and vision. Five participants were excluded due to poor performance, with accuracy close to random selection (25$\%$). To avoid bias, they were not informed of the study’s true objectives or the potential audio offset; instead, they were simply instructed to perform as accurately as possible.

\subsubsection{Results}

\begin{figure}[t]
    \centering
    \includegraphics[width=\columnwidth]{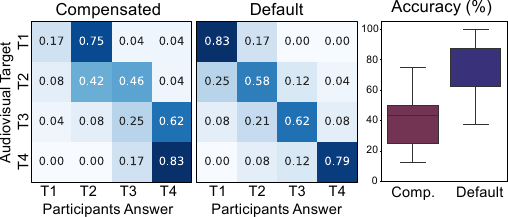}
    \caption{\textbf{Left:} Accuracy confusion matrix for target selection in conditions with compensation and without compensation (default). Higher diagonal accuracy indicates improved identification when the compensatory offset is not applied. \textbf{Right}: Overall participant accuracy in both conditions, with and without compensatory offset. Performance dropped significantly with the audio offset, indicating that the perceptual disparity does not arise without self-movement.}
	\label{fig:targetMovementConfusion}
\end{figure}

As shown in Figure~\ref{fig:targetMovementConfusion}, participants performed significantly better in the default condition ($70.8\%$) than in the compensated condition ($41.7\%$). The confusion matrix for the default condition exhibits a strong diagonal structure, indicating high accuracy and near-symmetry. Unlike the previous archery experiment, where self-movement led participants to favor the target in the direction of head rotation when making errors, here they selected both adjacent targets equally. This suggests that perceptual disparity does not occur in the absence of self-movement. Further supporting this, the compensated condition shows a diagonal shift toward the audio offset, indicating that the offset systematically biased participants toward the rightward target \review{(see supplementary material for further details)}.
The difference between accuracy distributions satisfies the normality assumption (Shapiro-Wilk test, w$=$0.96, p$=$0.78). The Student’s t-test revealed that participant performance when compensating for the perceptual offset ($\mu$=41.7$\%$, SE$=$0.050) significantly dropped compared to the default condition ($\mu$=70.8$\%$, SE$=$0.054), t(11)$=$-4.311, p$<$0.05, r$=$0.79. 

These results suggest that perceptual disparity does not emerge in the absence of self-movement and underscore the strong influence of the proprioceptive system on audiovisual localization. Therefore, attempting to compensate for the disparity hinders the user's performance, since perceptual disparity is no longer present. While Bernal-Berdun et al.~\cite{Bernal-Berdun2024spatial} attributed the perceptual disparity to the relative head-stimuli velocity, we found it only occurs when the velocity results from the user's own movement. 

\section{Discussion and Limitations}

\review{We have first examined the effects of audiovisual \textit{\dispT{}}, defined as a genuine misalignment between the spatial position of the auditory and visual cues of an audiovisual stimulus.}

\review{Our findings indicate that when participants concentrated exclusively on detecting mismatches between auditory and visual cues, they demonstrated high precision in identifying true disparities, particularly when misalignment exceeded $5^{\circ}$, which corresponds to the accuracy threshold of the generic HRTF used. This could suggest an important consideration when designing VR content, since even a slight misalignment could be detected and potentially impact the VR experience. However, our second experiment (Section~\ref{sec:task_demand}), which involved a more realistic VR scenario where participants were engaged in a demanding task beyond simply detecting the true disparity, suggests that users’ high sensitivity to misalignment may not pose a problem if the experience is sufficiently engaging. Although the limited number of participants and the use of only one type of headset may constrain the generalizability of our thresholds, we found that misalignments below $26.34^{\circ}$ to the left and $10.33^{\circ}$ to the right are unlikely to be perceived by the users. As such, these levels of disparity may be considered acceptable in realistic and immersive VR experiences.}

\review{Additionally, we have examined the impact of the spatial \textit{perceptual disparity} reported by previous work~\cite{Bernal-Berdun2024spatial} (Section~\ref{sec:perceptual_disp}).
We first examined whether the lack of rich audiovisual stimuli contributes to perceptual disparity (Section~\ref{sec:environmental}), interfering with a correct spatial localization. We replicated the experiment in such previous work and found the same perceptual disparity, evidencing its robustness to changes in VR environments. 
Participants tended to identify the sound associated with a visual stimulus to its left when rotating their head rightwards, despite it being co-located.
Introducing environmental audiovisual cues slightly improved overall performance, although not significantly. Some participants who performed worse found the radio sound distracting, while those who improved used it to infer the target’s position relative to the source. This suggests that carefully designed stimuli to avoid distraction may enhance spatial localization, though further studies are needed. Our small sample size and limited number of trials may have masked a meaningful effect, but even if present, such cues are unlikely to fully eliminate the perceptual disparity, since such a strong effect would likely have emerged despite our sample size. 
Unlike the previous work approach, which actively shifts audio cues to compensate for perceptual disparity, enriching the audiovisual environment may reduce disparity without altering the VR sound field. Proper integration of their approach requires aligning the audio when users are static and dynamically shifting it as they rotate. This implementation can be complex and, as highlighted in Section~\ref{sec:selfMovement}, improper application can significantly impair performance. In contrast, environmental cues could offer a more practical solution for VR, where multiple sound sources are common and audio manipulation may be technically challenging.}

\review{Finally, we have investigated whether the perceptual disparity arises from relative velocity between the head and the stimulus, as previously suggested, or specifically from self-initiated head rotation. To explore this, we have extended prior work by replicating the experimental conditions while removing the self-motion component, rotating the targets around the participant instead (Section~\ref{sec:selfMovement}).
Surprisingly, participants were able to accurately localize the audiovisual target in this condition, indicating that the perceptual disparity does not occur in the absence of self-rotation. This highlights the importance of considering proprioceptive cues in VR experiences. Additionally, we found that applying a compensatory approach, as proposed in the previous work, significantly impaired performance when no disparity was present. 
Implementing such a compensatory approach in real VR experiences seems an effective way to eliminate perceptual disparity. However, it must be carefully designed to be applied only during self-initiated rotation. Alternatively, VR experiences could be designed to minimize the need for self-rotation during localization tasks. Although this may reduce immersion and user freedom, it has been shown to effectively improve performance.}

\section{Conclusions}

We have investigated the impact of true and perceptual audiovisual disparities in complex, realistic VR applications. We found that while true disparities are less noticeable in engaging environments involving demanding tasks, perceptual disparities persist even in rich audiovisual settings. Notably, we found that auditory mislocalization due to relative motion between the stimulus and the participant's head only occurs during self-rotation.
%

In future work, we aim to further explore the robustness of these effects across diverse environments and stimuli, as well as \review{increase the number of participants and expand their diversity}. While we focused on a well-measured perceptual disparity with publicly available data, studying additional disparities in realistic VR settings remains an exciting venue to investigate.
We hope our findings contribute to the development of more immersive and coherent multimodal experiences.

\acknowledgments{
We would like to thank the members of the Graphics and Imaging Lab, the anonymous reviewers for their insightful comments, and the participants in the experiment.
This work has been supported by grant PID2022-141539NB-I00, funded by MICIU$\/$AEI$\/$0.13039$\/$501100011033 and by ERDF, EU, and by the Government of Aragon's Departamento de Educación, Ciencia y Universidades through the project “HUMAN-VR: Development of a Computational Model for Virtual Reality Perception” (PROY\_T25\_24). Additionally, Edurne Bernal-Berdun was supported by a Gobierno de Aragon predoctoral grant program (2021–2025).
The teaser has been designed using resources from Flaticon.com.}

\bibliographystyle{abbrv-doi}

\bibliography{main}
\end{document}